\documentclass[prl,aps,twocolumn]{revtex4}
\usepackage{amssymb}
\usepackage{amsmath}
\usepackage{epsfig}
\usepackage{t1enc}
\usepackage[utf8]{inputenc}

\begin{document}

\title{Debunking the black hole information paradox}

\author{Andrzej Dragan}

\email{dragan@fuw.edu.pl}

\affiliation{Institute of Theoretical Physics, University of
Warsaw, Ho\.{z}a 69, 00-681 Warsaw, Poland}

\begin{abstract}
The vivid debate concerning the paradox of information being lost when objects are swallowed by a black hole is shown to be void. We argue that no information is ever missing for any observer neither located above, nor falling beneath the event horizon. The information is preserved in a classical scenario of eternal black holes and semi-classical one allowing Hawking radiation.
\end{abstract}

\maketitle

Quantum theory of information is facing a crisis. At least these are claims of some authors referring to the paradox of information allegedly getting lost in the abyss of a black hole \cite{Giddings}. This fear has been seeded by Hawking \cite{Hawking} and the problem has its source in the popular view that any object can dive under the event horizon of a black hole. Therefore it is usually assumed that a static black hole is an example of an ideal black box that can absorb anything we want it to. But is such assumption legitimate? It is well known that a collapsing star forms its horizon asymptotically in time, so this process does not take place within a finite time \cite{Oppenheimer}. But even if we are given a formed black hole then due to the same reasons, a falling object as seen by an outside observer never reaches the horizon. In this Letter we consider both the point of view of a fiducial observer sitting outside the event horizon, and the point of view of an observer freely falling onto it. We discuss the scenario of a classical, eternal black hole and a scenario of a mortal black hole evaporating within a finite time \cite{Hawking2}. We prove that the paradox does not appear for neither of them.

\begin{figure}
\begin{center}
\epsfig{file=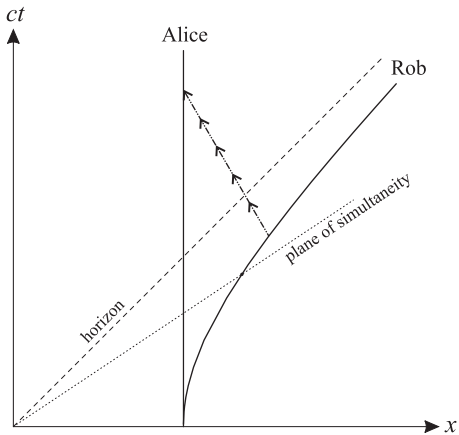, width=6cm} \caption{\label{jarekisloik} \sf \footnotesize Solid world-lines of accelerated Rob and free-falling Alice. Rob's event horizon is depicted with a dashed line, while the dotted line represents his plane of simultaneity. An arrowed line symbols a signal sent from Rob to Alice cloaked in horizon.}
\end{center}
\end{figure}

Consider Rindler Bob {\em alias} Rob moving with a uniform acceleration $a$ across flat spacetime and describing it within his noninertial frame of reference with the coordinates $(c\tau, \chi, y, z)$, where $\tau$ is the proper time of his clock, and $\chi$ measures the distance from the event horizon - Fig.~\ref{jarekisloik}. Rob, as a fiducial observer, is maintained at a fixed distance $\frac{c^2}{a}$ (in his frame of reference) from the horizon. Let Alice leave Rob at some point and move freely towards the event horizon - Fig.~\ref{jarekisloik}. Alice is an inertial observer and the relation between her coordinates $(ct, x, y, z)$, and Rob's noninertial coordinates is given by the Rindler transformation \cite{Wald}:
\begin{eqnarray}
\label{eq-RindlerTransform}
c\tau &=& \frac{c^2}{a_0}\,\mbox{atanh}\left(\frac{ct}{x}\right) \nonumber \\ \nonumber \\
\chi &=& \sqrt{x^2-c^2 t^2},
\end{eqnarray}
leading to the metric describing Rob's noninertial frame:
\begin{equation}
\label{eq-AliceMetric}
\mbox{d}s^2 = \frac{a^2\chi^2}{c^2} \mbox{d}\tau^2 - \mbox{d}\chi^2-\mbox{d}y^2-\mbox{d}z^2.
\end{equation}
It follows that a radius vector pointing at Rob's current position in spacetime determines his plane of simultaneity $\mbox{d}\tau = 0$ coinciding with the respective plane of simultaneity of the inertial observer temporarily co-moving with Rob - Fig.~\ref{jarekisloik}. Therefore an event on Rob's world line can only be simultaneous with an event on the Alice's world line if she has not crossed the horizon yet. It means that from the Rob's point of view Alice never reaches the horizon, she will only approach it asymptotically. Using the Rindler transformation \eqref{eq-RindlerTransform} one can derive Alice's equation of motion in the Rob's frame. Substituting the equation of the Alice's world line $x=\frac{c^2}{a}$ one gets:
\begin{equation}
\label{eq-RobMotion}
\chi(\tau) = \frac{c^2/g}{\cosh\left(a\tau/c\right)}>0,
\end{equation}
which confirms what we just said - Alice's distance $\chi$ from the horizon decreases asymptotically, but the horizon is never being crossed, although Alice claims that she crosses the horizon within finite proper time of hers, $t=\frac{c}{a}$. Let us underline that this discrepancy should {\em not} be related to an apparent effect nor optical illusion related to finite time of travel of light coming from Alice to Rob. It is rather, to quote Landau, {\em the extreme example of the relativity of time} \cite{Landau}. It can be concluded in the considered scenario that Rob does not suffer from the loss of information - Alice is always in Rob's domain of spacetime, therefore there is no information paradox. During Rob's lifetime Alice can always turn her rocket engines on and go back to Rob bringing back all the information she carried away.

Let us now investigate the same situation from the point of view of Alice falling freely onto the event horizon. According to her, the moment of crossing the border is not particularly interesting since nothing unusual happens then. The only disadvantage of crossing the horizon is that no signal sent by Alice can reach Rob ever since, she can only receive signals from him, as shown in Fig.~\ref{jarekisloik}. Therefore we are dealing with a fundamentally one-way communication channel. Since Alice is perfectly aware of how is Rob (he writes letters), she looses no information whatsoever about him. Again, there is no information paradox from this point of view. 

We will now study the event horizon surrounding a static black hole described by the Schwarzschild metric \cite{Wald}:
\begin{eqnarray}
\label{eq-SchwarzschildMetric}
\mbox{d}s^2 &=&  \left(1-\frac{R}{r}\right)c^2\mbox{d}\tau^2 - \frac{1}{1-R/r}\mbox{d}r^2
\nonumber \\ \nonumber \\
& &-r^2\left(\mbox{d}\theta^2 + \sin^2\theta\mbox{d}\phi^2\right),
\end{eqnarray}
where $R=\frac{2GM}{c^2}$ is the Schwarzschild radius and $r$ is defined so that $4\pi r^2$ is equal to the surface of a corresponding sphere centered in the origin of the coordinate system. Let us simplify the metric \eqref{eq-SchwarzschildMetric} in the proximity of the Schwarzschild radius, $r \approx R$. Introducing a new variable $\Delta r = r-R$, takes the metric to the form:
\begin{eqnarray}
\mbox{d}s^2 &=& \frac{\Delta r}{\Delta r +R}c^2\mbox{d}\tau^2 - \frac{\Delta r+R}{\Delta r}\mbox{d}\Delta r^2
\nonumber \\ \nonumber \\
& &-(\Delta r + R)^2\left(\mbox{d}\theta^2 + \sin^2\theta\mbox{d}\phi^2\right).
\end{eqnarray}
However, the spatial coordinate $\Delta r$ does not measure the real distance, because $r^2$ has been defined as the surface of the sphere divided by $4\pi$. The real, infinitesimal distance near the horizon ($\Delta r \ll R$) is given by the expression $\mbox{d}\chi = \sqrt{-g_{rr}(r)}\mbox{d}r \approx \sqrt{\frac{R}{\Delta r}} \mbox{d}\Delta r = 2\mbox{d}\sqrt{R\Delta r}$. We also change the other spatial coordinates according to the following substitution: $R^2\left(\mbox{d}\theta^2 + \sin^2\theta\mbox{d}\phi^2\right)\equiv\mbox{d}y^2 + \mbox{d}z^2$ obtaining the new form of the Schwarzschild metric in the direct proximity of the event horizon:
\begin{equation}
\mbox{d}s^2 = \frac{\chi^2c^2}{4R^2}\mbox{d}\tau^2 - \mbox{d}\chi^2 - \mbox{d}y^2 - \mbox{d}z^2,
\end{equation}
which is no different than the metric \eqref{eq-AliceMetric} of the Rob's noninertial frame moving with the uniform acceleration $a=\frac{c^2}{2R}$. It doesn't come as a surprise that Unruh radiation \cite{Unruh} coincides with the Hawking radiation near the black hole's horizon. What is more important for our discussion is that the conclusions we have drawn considering Rob's noninertial reference frame apply directly to the case of a fiducial observer in the proximity of a static black hole attracting Alice gravitationally. Therefore also in this case none of the observers is troubled by the information paradox. Brutally speaking: from our point of view never has any object been swallowed by an event horizon of no black hole. In the proximity of the event horizon there is a graveyard of objects that have almost reached it. Not to mention the fact that the process of formation of the horizon itself during the gravitational collapse of a star is also only asymptotic \cite{Oppenheimer}. 

\begin{figure}
\begin{center}
\epsfig{file=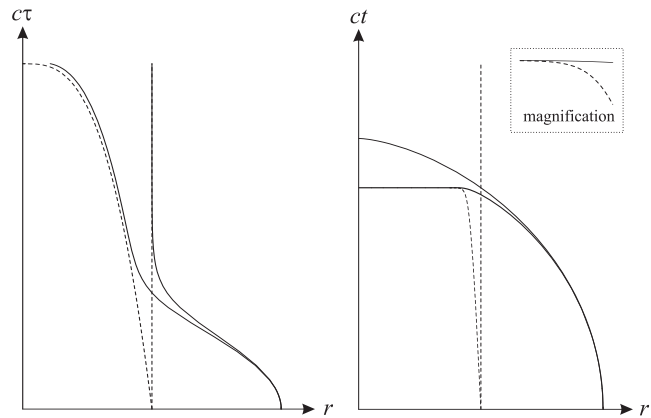, width=8.5cm} \caption{\label{spadek} \sf \footnotesize Alice freely falling onto the eternal and mortal black holes (solid lines) in Rob's reference frame (left) and in her own reference frame (right). Dashed lines represent respective event horizons for eternal and evaporating black holes. According to Alice, the horizon can only be crossed in the absence of the Hawking radiation - see magnification of the evaporating case (also rescaled).}
\end{center}
\end{figure}

The above reasoning is related to the so-called black hole complementarity \cite{Susskind} and applies to the case of eternal black holes in the absence of the Hawking radiation. But what happens if we take into account the black hole evaporation process \cite{Hawking2}? According to Susskind \cite{Susskind} this semi-classical scenario might lead to the possibility of information cloning, which is forbidden by the quantum theory \cite{Zurek} and only subtle arguments of the string theory can save the day. In the following we argue that no string theory is needed to debunk the paradox. Let us suppose that the black hole has been already formed somehow and Alice is heading towards its horizon. Due to the Hawking radiation the black hole evaporates within finite time \cite{Hawking2} and therefore according to Rob, Alice can only reach the horizon exactly the moment the black hole disappears. This seems to indicate that in the Alice's frame of reference, by the time she reaches the horizon there will be no black hole left to visit. This hand-waving argument seems to show that the black hole radiation process makes the information paradox even easier to debunk. Before we have a closer look at this, let us underline that the ultimate nature of the Hawking radiation is not fully known and some authors speculate that a small remnant of the singularity may survive the evaporation on the Planckian scale \cite{Oppenheim}. But even if this is the case, such remnant is way too small to have any impact onto Alice. It would be more appropriate to say that Alice absorbs such a microscopic black hole skeleton than the opposite. 

To give our arguments some quantitative support, let us study the equation of motion of Alice radially falling onto the black hole. Although for the eternal black hole the equation can easily be derived analytically, the case of the evaporating black hole needs numerical treatment. One can consider a simplistic toy-model of a quasi-static evaporation \cite{Aste}, for which the Schwarzschild radius in \eqref{eq-SchwarzschildMetric} shrinks in time according to the Hawking's formula $R(\tau) = (R(0)-k\tau)^{\frac{1}{3}}$, where the constant $k$ depends on the initial Schwarzschild radius, and the evaporated particles do not influence the metric. The resulting trajectories are depicted in Fig.~\ref{spadek}, where on the left we present Rob's point of view defined by $r(\tau)$ and on the right we show Alice's point of view $r(t)$, where $r$ defines Alice's position, $\tau$ is Rob's temporal coordinate defined by the Schwarzschild metric \eqref{eq-SchwarzschildMetric} and $t$ is Alice's proper time. For the eternal black hole the results are well-known: according to Rob, Alice (solid line in Fig.~\ref{spadek}) never reaches the horizon (dashed line), while according to Alice, she crosses the horizon and reaches the singularity within a finite proper time $t$. When the black hole evaporates, Rob still claims that Alice never reaches the horizon, but Alice's point of view changes dramatically - Fig.~\ref{spadek}. According to her, the horizon is rapidly sucked into the singularity, as she gets closer, so that she can only touch it when the black hole eventually disappears. She never crosses the horizon in contrast with the fully-classical case. Since Alice does not cross the horizon, no contradiction with no-clonning theorem arises and the information paradox turns out to be void again.

Another question is whether the free-falling Alice records Hawking radiation. Suppose the answer is positive. Since the flux of radiated particles, which is proportional to $\frac{\mbox{\scriptsize d}R}{\mbox{\scriptsize d}t}$, blows up to infinity when Alice gets closer and closer to the horizon, it might be somehow problematic for her to survive the encounter with the event horizon. However such possibility would be in conflict with the principle of equivalence. Clearly, free-moving Alice in flat spacetime depicted in the Fig.~\ref{jarekisloik} does not perceive Unruh radiation and one might expect the same from the free-falling observer approaching the black hole. But if we assume that the Hawking radiation is invisible to Alice, it is hard to explain for Alice the sudden disappearance of the black hole. Probably the definite answer to these issues can only be given with the use of the still unknown quantum theory of gravity, but whatever the answer is, for now it is definitely too early to herald the fall of the quantum theory of information.

To summarize our discussion, we have argued that no object can be consumed by the event horizon of a black hole, so there is no contribution to the Hawking radiation from the matter falling onto the horizon. The radiation may be fueled only by the matter bound within the Schwarzschild radius from the very beginning of the evolution. On the ground of the simplistic toy-model of evaporation we also concluded that all the matter trapped near the horizon is released within finite time. However our result clearly does not depend on the particular model of evaporation, so for arbitrary decay curve of the black hole qualitative conclusions remain the same.

We would like to thank Iwo Białynicki-Birula and Krzysztof Pachucki for very interesting sparring discussions.

\end{document}